\documentclass[conference]{IEEEtran}
\IEEEoverridecommandlockouts
\usepackage{cite}
\usepackage{amsmath,amssymb,amsfonts}
\usepackage{algorithmic}
\usepackage{graphicx}
\usepackage{textcomp}
\usepackage[table,xcdraw]{xcolor}
\usepackage{tikz}
\usepackage{url}
\usetikzlibrary{calc,trees,positioning,arrows,chains,shapes.geometric,%
    decorations.pathreplacing,decorations.pathmorphing,shapes,%
    matrix,shapes.symbols}
\newlength{\mytw}
\def\BibTeX{{\rm B\kern-.05em{\sc i\kern-.025em b}\kern-.08em
    T\kern-.1667em\lower.7ex\hbox{E}\kern-.125emX}}
\tikzset{
>=stealth',
  punktchain/.style={
    rectangle, 
    rounded corners, 
    draw=black, very thick,
    text width=10em, 
    minimum height=3em, 
    text centered, 
    on chain},
  line/.style={draw, thick, <-},
  element/.style={
    tape,
    top color=white,
    bottom color=blue!50!black!60!,
    minimum width=8em,
    draw=blue!40!black!90, very thick,
    text width=10em, 
    minimum height=3.5em, 
    text centered, 
    on chain},
  every join/.style={->, thick,shorten >=1pt},
  decoration={brace},
  tuborg/.style={decorate},
  tubnode/.style={midway, right=2pt},
}
\begin{document}

\title{A Systematic Mapping Study on Blockchain Technology for Digital Protection of Communication with Industrial Control
}
\author{\IEEEauthorblockN{Kirill Loisha}
\IEEEauthorblockA{\textit{Department of Informatics and Mathematics} \\
	\textit{HTW Berlin University of Applied Sciences, Germany}\\
	Berlin, Germany \\
	kiril.loisha@hhtw-berlin.de}

\and
\IEEEauthorblockN{Javad Ghofrani} 
\IEEEauthorblockA{\textit{Department of Informatics and Mathematics} \\
\textit{HTW Dresden University of Applied Sciences, Germany}\\
Dresden, Germany \\
javad.ghofrani@gmail.com}
\and

\IEEEauthorblockN{ Dirk Reichelt}
\IEEEauthorblockA{\textit{Department of Informatics and Mathematics} \\
\textit{HTW Dresden University of Applied Sciences, Germany}\\
Dresden, Germany \\
email address or ORCID}

}
\maketitle

\begin{abstract}
In the next few years, Blockchain will play a central role in IoT as a technology. It enables the traceability of processes between multiple parties independent of a central instance. Blockchain allows to make the processes more transparent, cheaper, and safer. This research paper was conducted as systematic literature search. Our aim is to understand current state of implementation  in context of Blockchain Technology for digital protection of communication in industrial cyber-physical systems. We have extracted 28 primary papers from scientific databases and classified into different categories using visualizations. The results show that the focus in around 14\% papers is on solution proposal and implementation of use cases "Secure transfer of order data" using Ethereum Blockchain, 7\% papers applying Hyperledger Fabric and Multichain. The majority of research (around 43\%) is focusing on solution development for supply chain and process traceability.
\end{abstract}


\section{Introduction}
The success of the manufacturing industry based on the revolutionary innovations in various technologies in hardware and software \cite{8666558}. The conception of the Internet of Things (IoT) has already become a part of modern factories for automatizing of large-scale processes~\cite{Microsoft2019}. 

Manufacturing sector brings many different stakeholders together. This industry has mostly common fraud problems comparing to any sector. The significant rate (56\%) of respondents reported the issues related to vendor or
procurement fraud~\cite{kroll}. For some companies developing know-how products, intellectual property infringement could be fatal for the whole business. Not surprisingly, these issues make companies look for new ways to avoid them~\cite{deloitte2014}.

Blockchain technology, which stays behind Bitcoin, is nowadays a hype technology\cite{8704309}. Its development could be revolutionized like the appearance of the Internet at the beginning of 90th's. Blockchain provides a secure mechanism for achieving an independent trusted between business partners, excluding intermediaries. 
The rest of the paper is organized as follows. Section \ref{sec:research:method} discusses information about research methods. In Section \ref{sec:results} illustrates the results of searching and screening for relevant papers for this research, while section \ref{sec:analysis} answers the research questions using these results. Section \ref{sec:conlcusion} concludes the paper.


\subsection{Overview of Blockchain}\label{sec:background}
Blockchain is a new and successful combination of existing technologies. The technical concept of Blockchain describes how data are distributed saved across the user systems using cryptography algorithm.
Blockchain organized logically centralized and organizationally decentralized. As a result, the Blockchain represents a distributed database that maintains an ever-expanding list of decentralized transaction, events, or records in hash form. The data is maintained in a distributed register (Distributed Ledger Technology) and all participants have a copy of the entire register. In this distributed approach, the data is grouped into individual blocks that are linked together to ensure the chronological order and immutable data integrity of the entire data set~\cite{8715005}. 

The innovation of Blockchain technology is that existing approaches have been successfully put together. Following approaches are the main components of Blockchain:
\begin{itemize}
\item \textbf{Peer-to-peer network:} In this peer-to-peer network (P2P network), communication runs without a central point. All participants or nodes are connected to each other and communicate with each other at the same level. Since the nodes are equal to each other, or can use services and make them available at the same time, there is no classic client-server structure\cite{8704309}.
\item \textbf{Cryptography:} With the help of methods from cryptography, the distributed register is protected against manipulation and abuse. This enables traceability, data integrity and authentication of the data source\cite{LI2018133}.
\item \textbf{Consensus mechanism:} The consensus mechanism defines the criteria that provide evidence of permission to create new blocks (mining). To reach a consensus, various consensus algorithms have been developed\cite{Innerbichler:2018:FBA:3211933.3211953}.
\end{itemize}

Due to the different uses of Blockchain technology, there are different variations on how the Blockchain is constructed.
In a \textit{public Blockchain}, there are no restrictions on who can see the public data and validate the transactions\cite{LI2018133}. Furthermore, the Blockchain data may be encrypted and understandable only to the authorized user. In the case of a \textit{private Blockchain}, the completed consortium of those who access the Blockchain and are allowed to validate transactions is predefined~\cite{8560153}. In a \textit{permissionless Blockchain}, there are no restrictions on the identity of the participants who are allowed to conduct the transactions. In a \textit{permissioned Blockchain}, the user group that can execute the transactions and generate new blocks is predefined and known~\cite{8678753}.
\subsection{Smart Contracts}
In addition to a decentralized database system for transactions, Blockchain technology is also a platform for the automation of processes, regulations and organizational principles.
Smart contracts are new, intelligent forms of contracts. These are to be understood as peer-to-peer applications, which are distributed with the underlying Blockchain technology, such as Bitcoin, Ethereum or Hyperledger. Hyperledger also uses the term "chaincode"~\cite{Wang:2019:SSB:3302505.3310086}.

The smart contracts enable the determination of the conditions that lead to certain decisions by the data provided; the automatic processing of contracts; the permanent and real-time monitoring of contract terms and the automatic enforcement of the rights of contractors. The smart contracts provide not only the information about the Blockchain network and the distribution of data, but also the business logic~\cite{Baumung2018135}.

These P2P applications can be programmed, stored in the Blockchain and executed in there. Therefore, they have the same advantages as the Blockchain itself. In Bitcoin, the smart contracts are created in the form of scripts. In order to simplify the development of smart contracts for Ethereum platform, a new specific programming language "Solidity" has been developed~\cite{Baumung2019456}. Compared to the scripting language in Bitcoin, where many program constructs, such as loops, are missing, Solidity is a higher-level, more abstract language that resembles the JavaScript language. The chain codes are written in various high-level languages, such as Java or Go, and during execution, access is made to the data stored in the Blockchain, or to read out the existing information and store new ones. These scripts are stored in the Blockchain at a particular address. This address is determined when the integration of  smart contract into the Blockchain is decided. When an event prescribed in the contract has occurred, a transaction is sent to this address. The distributed virtual machine executes the code of the script using the data sent with the transaction.
\subsection{State of Research on Blockchain}
Yli-Huumo et. al~\cite{Yli_Huumo_2016} aim to understand the current research state of Blockchain technology, its technical challenges and limitations. This systematic review illustrates an sharply increasing number of publications each year beginning from 2012. It shows a growing interest in Blockchain technology. 
Swan~\cite{swan2015blockchain} identified seven technical challenges of Blockchain for the future. Modern Blockchain implementations have to ensure security, throughput, size and bandwidth, performance, usability, data integrity and scalability. Being public Blockchains the \textit{throughput} in the Bitcoin and Ethereum networks is from 10tps to 100tps (transactions per second).  For example, VISA Payment System proceeds 2,000tps. But a permissioned Blockchain Hyperledger Fabric overcomes these challenges~\cite{Alharby2017}. In order to achieve adequate security in the Bitcoin network validation of transaction  takes roughly 10 minutes \textit{(latency)}. In February 2016 the size of Bitcoin register was 50,000 MB. Current \textit{size} of Bitcoin is 1 MB. This is the serious limitation of \textit{bandwidths} for Blockchain, which should be solved to increase amount of transaction handled by register. The 51\% attach on Blockchain network is still significant security issue. If the majority of the network  will be controlled by hackers, it will be possible to manipulate Blockchain. Issue of \textit{waster resources} is caused  by Proof-of-Work effort in the mining process mainly in Bitcoin, which required huge amounts of energy.  But there are other consensus algorithms, like Proof-Of-Stake, which are energy friendly. \textit{Usability} problems resulting  from difficulty of using Bitcoin API~\cite{Yli_Huumo_2016}. \textit{Versioning, hard forks, multiple chains} refers to a small chain with a small number of nodes, where a possibility of  51\% attach is higher. Another issue become possible  when chains are split for administrative or versioning purposes.
\begin{figure*}[htbp]
\centerline{\includegraphics[width=\textwidth,height=4.5cm]{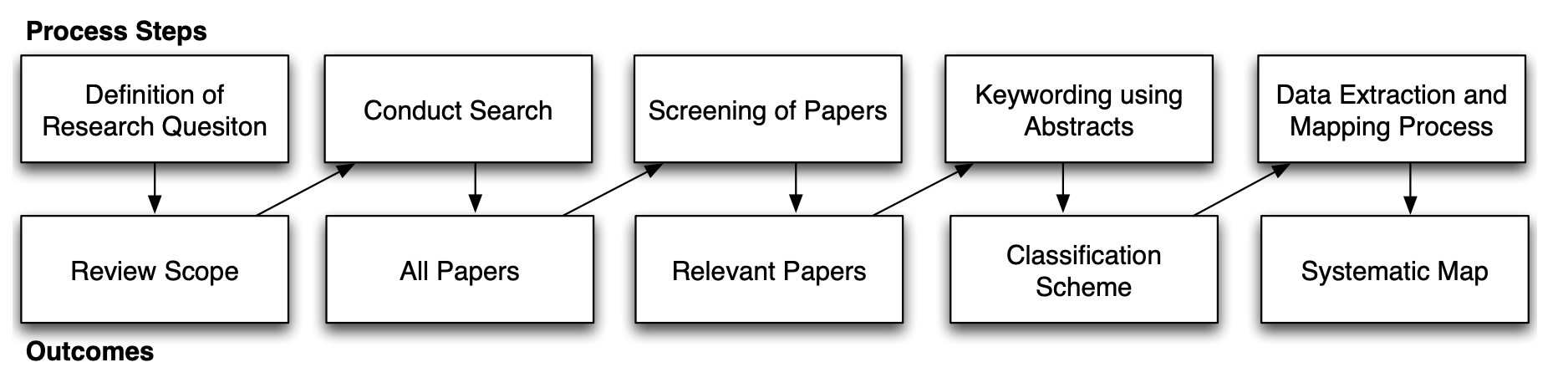}}
\caption{The Systematic Mapping Process~\cite{Petersen:2008:SMS:2227115.2227123}.}
\label{fig:001}
\end{figure*}
\section{Research method}\label{sec:research:method}
A systematic mapping study was selected to identify and classify primary studies to provide a systematic overview on the topics of industrial manufacturing and Blockchain. Petersen et al.~\cite{Petersen:2008:SMS:2227115.2227123} presented the guidelines for systematic mapping study, which we followed to conduct this study.

The process for the systematic mapping study falls into a  five-phase process as depicted in Figure \ref{fig:001}: (1) Define research questions; (2) Search for primary studies; (3) Identify inclusion and exclusion criteria and screen primary studies based on these criteria; (4) Classify primary studies; (5) Mapping the data.
\subsection{Research questions}
The first step in systematic mapping study is the definition of the research questions. The purpose of this research is to classify current research and identify pertinent themes which relate directly to Blockchain technologies in manufacturing. This leads to the following research questions (RQs): \\

\textbf{RQ1:} \textit{What are the problems between stakeholders in the manufacturing
industry? } \\
Rationale: The intention of this question is to identify current gaps in relationships of interested parties in manufacturing industry.\\

\textbf{RQ2:} \textit{What are the data to secure in manufacturing process?} \\
Rationale: This question aims to identify the data, which should be insured during the whole process.\\

\textbf{RQ3:} \textit{What are the use cases of Blockchain technology for manufacturing industry?} \\
Rationale: The intention of this question is to figure out the possible usage of Blockchain in manufacturing area.\\

\textbf{RQ4:} \textit{What Blockchain frameworks are suitable for the scenario "Assignment of production orders to an external manufacturer"?} \\
Rationale: With this question, we investigate the existing Blockchain solutions to find appropriate framework. \\
\subsection{Search Strategy}
The search strategy is key to ensure a good starting point for the identification of studies and ultimately for the actual outcome of the study. An extensive and broad set of primary studies was needed to answer the research questions. The most popular academic databases in the domain of software engineering were selected to be used in this systematic mapping to search for potentially relevant papers:
\begin{itemize}
\item ACM Digital Library\footnote{http://dl.acm.org}
\item IEEE Xplore Digital Library\footnote{http://ieeexplore.ieee.org}
\item Scopus\footnote{https://www.scopus.com}
\item Science Direct\footnote{https://www.sciencedirect.com}
\end{itemize}

Finding possibly relevant publications to answer the research questions requires creating an appropriate search clause. We chose the terms "Blockchain" and "Manufacturing industry" for this study as the main search keyword core, it focuses on Blockchain technology, manufacturing, production processes.
The final search strings were extended with alternative synonyms for main keywords. The term "distributed ledger" is a basic technology for "blockchain". We considered papers mentioning distributed manufacturing, manufacturing execution, programmable logic controller" and included them into the search clause.
Regarding the keywords for the search, after some exploratory searches using different combination of keywords, the researchers jointly established the final string to be used in the search for papers in the databases. Search terms with similar meanings were grouped in the same group and combined using the OR logical operator. To perform automatic searches in the selected digital libraries, the AND logical operator were used between combined terms of different groups, depicted in Table \ref{tab:01}:
\begin{table}[htbp]
\centering
\label{tab:01}
\caption{Searches in databases.}
\begin{tabular}{|l|l|}
\hline
\rowcolor[HTML]{EFEFEF} 
\textbf{Database} & \textbf{Search}\\ 
\hline
ACM               & \begin{tabular}[c]{@{}l@{}}(+("Blockchain" "Distributed Ledger") +\\("Manufacturing Execution" "Programmable Logic \\ Controller" "Manufacturing" "Distributed \\manufacturing")\end{tabular}                                                             \\ \hline
IEEE              & \begin{tabular}[c]{@{}l@{}}('Blockchain' OR 'Distributed Ledger') AND \\ ('Industrial Control' OR 'Manufacturing Execution' \\ OR 'Programmable Logic Controller' OR \\'Manufacturing industry' OR 'Distributed \\manufacturing')\end{tabular}           \\ \hline
Scopus            & \begin{tabular}[c]{@{}l@{}}ALL (("Blockchain" OR "Distributed Ledger") AND\\ ("Industrial Control" OR "Manufacturing Execution" \\ OR "Programmable Logic Controller") OR \\ ("Manufacturing industry" OR "Distributed\\ manufacturing"))\end{tabular} \\ \hline
Science Direct    & \begin{tabular}[c]{@{}l@{}}("Blockchain" OR "Distributed Ledger") AND \\ ("Industrial Control" OR "Manufacturing Execution" \\ OR "Programmable Logic Controller" OR \\ "Manufacturing" OR "Distributed manufacturing")\end{tabular}                    \\ \hline
\end{tabular}
\end{table}

The search string was applied to title, abstract, full-text and keywords, and limited to journal papers written in English. The search was performed at the beginning of 2016. A total of 258 papers were retrieved from the different databases, which are focusing on research regarding information technologies, as on 21st May 2019, 10.30 am (CEST) and displayed in Table \ref{tab:studies}. 
\begin{table}[htbp]
\label{tab:studies}
\caption{Number of studies per database.}
\centering
\begin{tabular}{|l|l|l|l|l|}
\hline
\rowcolor[HTML]{EFEFEF} 
\textbf{Database} & \textbf{\begin{tabular}[c]{@{}l@{}}Search\\ results\end{tabular}} & \textbf{\begin{tabular}[c]{@{}l@{}}Final\\ results\end{tabular}} & \textbf{\begin{tabular}[c]{@{}l@{}}\% final papers\\ from search results\end{tabular}}  \\ \hline
ACM               & 24   & 4 & 1.6 \%         \\ \hline
IEEE              & 61   & 12 & 4.7 \%         \\ \hline
Scopus            & 159  & 6 &  2.3 \%       \\ \hline
Science Direct    & 14   & 6 & 2.3 \%         \\ \hline
\textbf{Total}    & \textbf{258} &  \textbf{28} &  \\ \hline
\end{tabular}
\end{table}

\subsection{Screening for relevant papers}
This step is to exclude all research papers that are irrelevant to the research questions. To accomplish this step, We followed the screening approach described by Yli-Huumo et al.~\cite{Yli_Huumo_2016} and defined inclusion/exclusion criteria.\\

\textbf{Inclusion criteria} 

To be considered for inclusion in the study, the research being evaluated had to originate from an academic source, such as a journal or conference, and clearly show its contribution was focused on applying Blockchain in manufacturing. Studies are accessible electronically. The paper title includes "Blockchain".\\

\textbf{Exclusion criteria}

\begin{figure}[htbp]
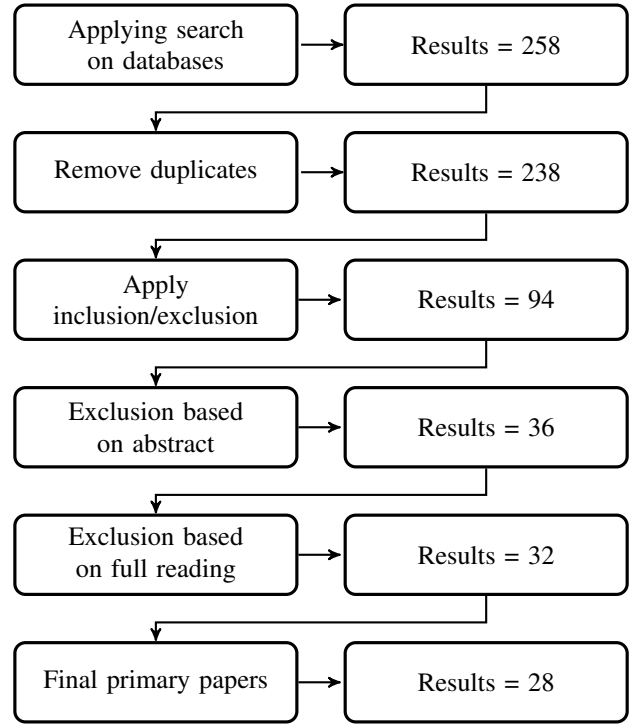

\caption{Search and Selection Process of the Papers.}
\label{fig:sum}
\end{figure}

For those publications that passed the inclusion criteria, two filters were applied to reduce the publications to only those that were deemed to be directly aligned with the focus of the study. These filters are described as follows:
\begin{itemize}
\item Study focuses on financial sector
\item Study contains "Bitcoin" or "Cryptocurrency"
\end{itemize}
It was decided to apply exclusion criteria on titles, keywords, abstracts and full-text. The final inclusion or exclusion could be decided based on the reading of publication’s full text. Thus, this phase only eliminates publications clearly not within this study’s scope and publications failing on formal requirements (such as duplicated papers). In detail, we identified 230 publications (about 90\% of all papers) outside this study’s scope: 20 duplicates, 156 papers accordingly with inclusion/exclusion criteria and other based on abstract reading.
The search process is summarized in Figure \ref{fig:sum}.\\

\begin{tikzpicture}
[node distance=.6cm,
start chain=going below,]
    \node (step0) [punktchain ]  {Applying search on databases};
        \begin{scope}[start branch=venstre,
            every join/.style={->, thick, shorten <=1pt}, ]
                \node[punktchain, on chain=going right, join=by {->}]
            (result0){Results = 258};
        \end{scope}
    \node (step1) [punktchain ]  {Remove duplicates};
        \begin{scope}[start branch=venstre,
            every join/.style={->, thick, shorten <=1pt}, ]
                \node[punktchain, on chain=going right, join=by {->}]
            (result1){Results = 238};
        \end{scope}
    \node[punktchain] (step2) {Apply inclusion/exclusion};
        \begin{scope}[start branch=hoejre1,]
            every join/.style={->, thick, shorten <=1pt}, ]
                \node [punktchain, on chain=going right, join=by {->}] 
            (result2) {Results = 94};
        \end{scope}
    \node[punktchain] (step3) {Exclusion based on abstract};
        \begin{scope}[start branch=hoejre1,]
            every join/.style={->, thick, shorten <=1pt}, ]
                \node [punktchain, on chain=going right, join=by {->}] 
            (result3) {Results = 36};
        \end{scope}
    \node[punktchain] (step4) {Exclusion based on full reading};
        \begin{scope}[start branch=hoejre1,]
            every join/.style={->, thick, shorten <=1pt}, ]
                \node [punktchain, on chain=going right, join=by {->}]
            (result4) {Results = 32};
        \end{scope}
    \node[punktchain] (step5) {Final primary papers};
        \begin{scope}[start branch=hoejre1,]
            every join/.style={->, thick, shorten <=1pt}, ]
                \node [punktchain, on chain=going right, join=by {->}]
            (result5) {Results = 28};
        \end{scope}
  \draw[|-,-|,->, thick,] (result0.south) |-+(0,-1em)-| (step1.north);
  \draw[|-,-|,->, thick,] (result1.south) |-+(0,-1em)-| (step2.north);
  \draw[|-,-|,->, thick,] (result2.south) |-+(0,-1em)-| (step3.north);
  \draw[|-,-|,->, thick,] (result3.south) |-+(0,-1em)-| (step4.north);
  \draw[|-,-|,->, thick,] (result4.south) |-+(0,-1em)-| (step5.north);
  \end{tikzpicture}
\subsection{Key-wording using Abstracts}
Yli-Huumo et al.~\cite{Yli_Huumo_2016} proposed the key-wording technique to classify all relevant research papers. We first went thought the abstract of each paper to identify the most important keywords and the key contribution. This purpose used to classify papers under different categories.  In some cases where it was difficult to classify a paper using
its abstract, we looked through its introduction and conclusion. After classifying all papers, we read the papers and made changes to the classification when necessary.
\subsection{Data extraction and mapping process}
The bubble plot was designed to collect all the information needed to address the research questions of this mapping study. During the process of data extraction, we recorded major terms to Excel, which helped me to generate categories and proceed quality analysis. These data items embrace the main goals of papers. We used both qualitative and quantitative synthesis methods.
\\

The final search process results 28 papers performed at the beginning of 2017. Their distribution over databases and percentage from all search results are illustrated in Table \ref{tab:studies}.\\
\section{Study Results}\label{sec:results}
This section demonstrates the findings from those the data was extracted regarding use cases of Blockchain in manufacturing industry, research type and attributes of Blockchain technology.
\begin{figure}[htbp]
\centerline{\includegraphics[width=4.25cm,height=4.25cm]{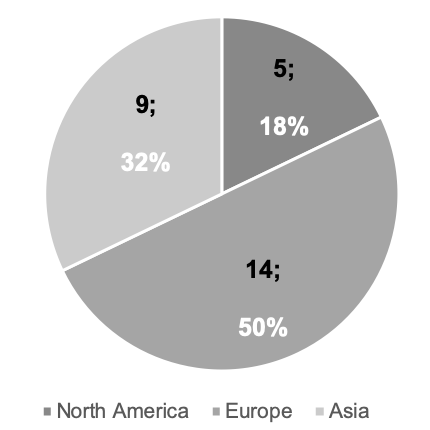}}
\caption{Continent wise distribution of studies.}
\label{fig:country}
\end{figure} 

\textbf{Top Countries and continents}

Geographic distribution of the selected primary papers is shown in Fig. \ref{fig:country}. The top continent was Europe with 14 studies being conducted there. Asia was second with 9 studies followed by America with 5 studies. China and Germany contributed towards 6 and 3 studies respectively. The rest of the countries had two or less papers published. It shows, that Blockchain technology has attracted attention worldwide.  \\

\textbf{Year Wise Distribution}

\begin{figure}[htbp]
\centerline{\includegraphics[width=6cm,height=4cm]{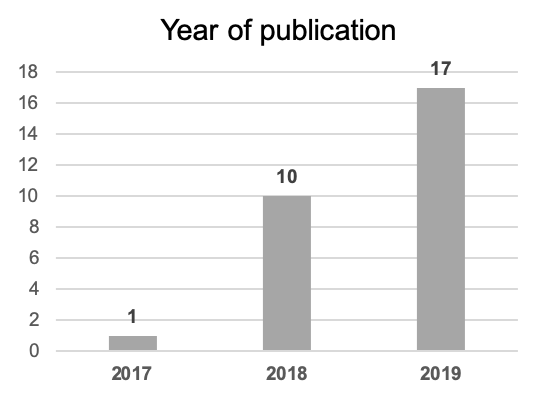}}
\caption{Publication year of the selected primary papers.}
\label{year}
\end{figure}

Figure \ref{year} depicts the distribution of the papers from 2017 to 2019 year. Literature related to Blockchain in manufacturing industry has increased enormously in the last 2 years. Due to increasing papers of applying Blockchain in industrial context it expected to save this trend for next years. \\

\textbf{Classification of research.}

All  of  the  publications  in  the  study were classified by considering the following
criteria: (1) Use case of Blockchain, (2) Research facet and (3) Blockchain facet.
\subsection{Use case of Blockchain:}
In order to answer RQ3, we classified  the  publications  in  the  study  under five dimensions. These dimensions describe different use cases of Blockchain in manufacturing industry on the current state of research in the area. The use cases of Blockchain in industrial manufacturing are follows: \\

\textbf{Secure transfer of order data.} This use case describes how the production orders can be assigned to an external manufacturer and securely transmitted between different systems. It enables mutual interaction between the producer and the customer~\cite{8250199}.\\

\textbf{Product data storing.} The data can be intercepted during transmission from the user's computer to the cloud systems. In these use cases the focus is on secured storing of products data in Blockchain.\\

\textbf{Supply chain, Process traceability.} Creating and distributing of goods can span over multiple  locations, hundreds of stages etc. The use case aims to provide the ability to trace process in supply chain from procurement of raw materials to production~\cite{8711819}.\\

\textbf{Prevention of fraud, Protection of Intellectual Property (IP).} The main point of this use case is to prove of products origin and intend for prevention of manipulation providing an indelible and traceable record of changes.\\

\textbf{Industrial IoT (IoT), Automation.} This use case illustrates how Blockchain can be used for integration with industrial IoT in automation context. \\

\begin{figure}[htbp]
\centerline{\includegraphics[width=9cm,height=5cm]{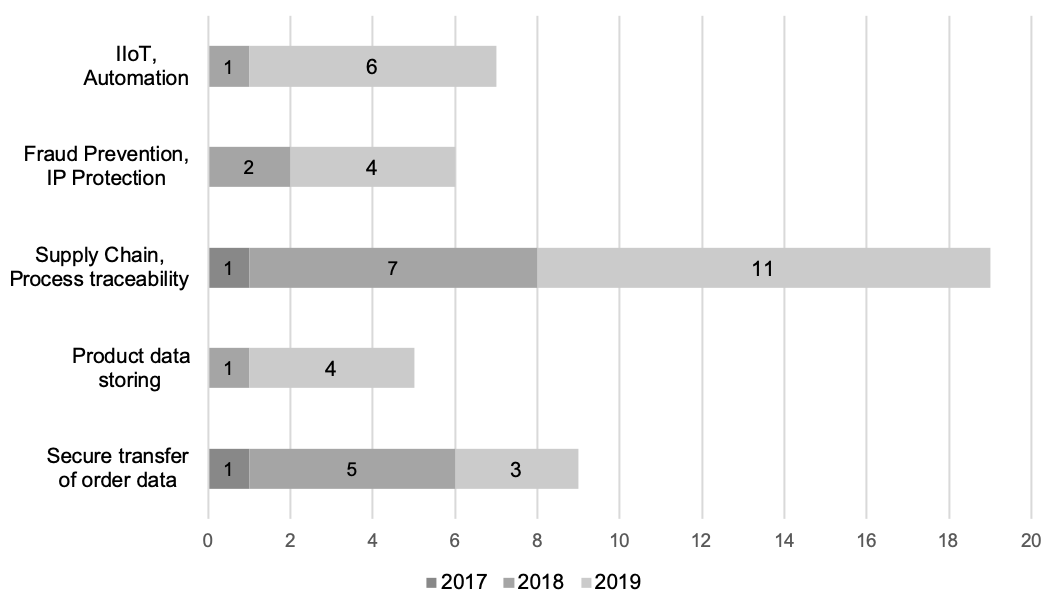}}
\caption{Publication year of the identified use cases.}
\label{trend}
\end{figure}

Figure \ref{trend} shows the amount of publications where each of use case above is described year wise. This figure illustrates, that the most researched topic is supply chain and process traceability (19 papers). Less researched (5 papers) is the scenario of how could the product data be stored in Blockchain. All use cases, except "Secure transfer of order data" show the trend of increasing interests. 

\subsection{Research Facet:}
This facet is inspired by~\cite{Petersen:2008:SMS:2227115.2227123} classify publications according to the type of research they contribute. (1) Review paper summarizes the current state of understanding on a topic. (2) Conceptual paper addresses a question that cannot be answered simply by getting more factual information. (3) Solution proposal includes an illustration or example of a solution to a particular problem. (4) Implementation research  provides a prototypical development of a solution. (5) Case Study provides an up-close, in-depth, and detailed examination of a subject of study.

\begin{figure*}[htbp]
\centerline{\includegraphics[width=\textwidth,height=11cm]{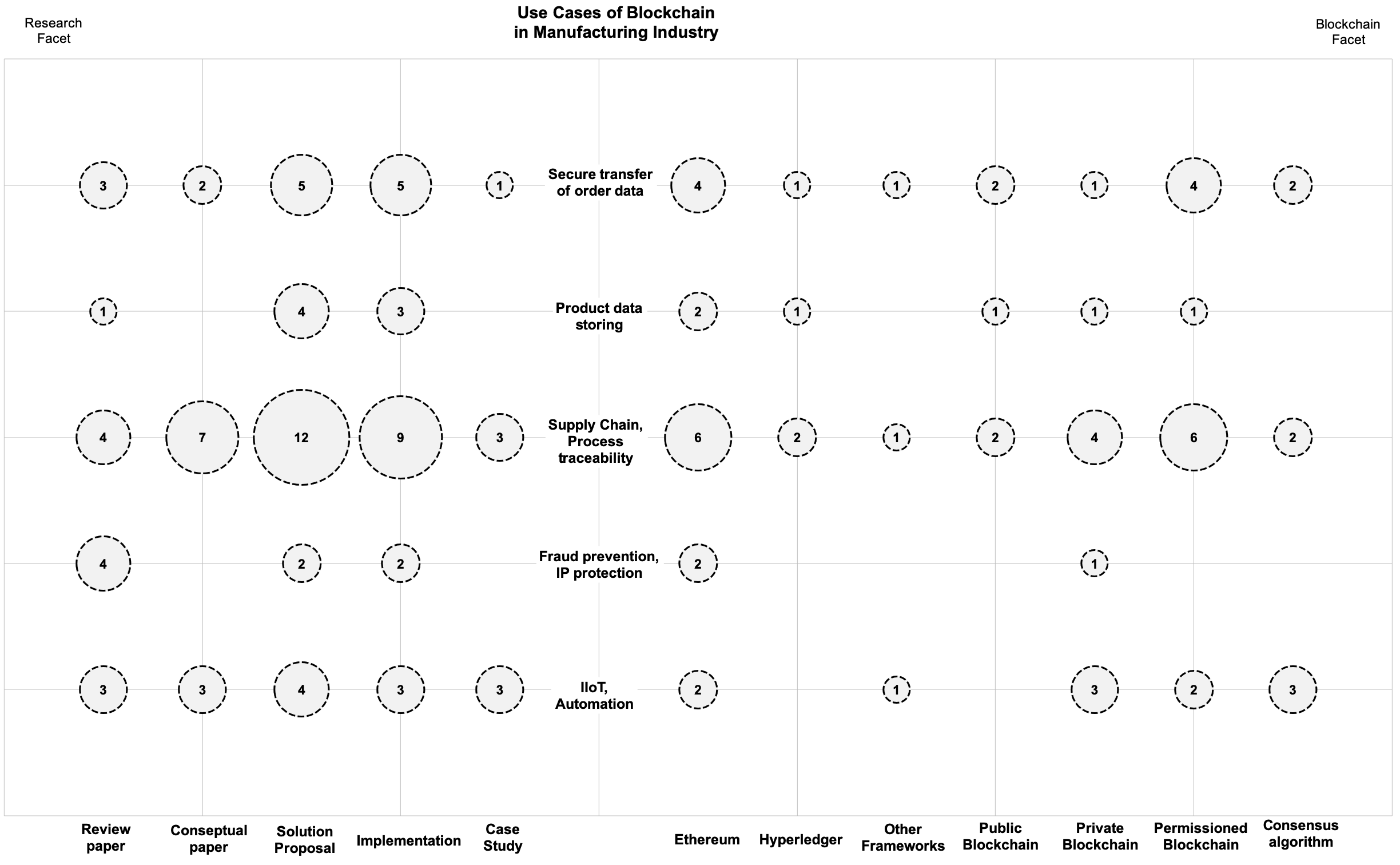}}
\caption{Visualisation of a Systematic Map in the Form of a Bubble Plot}
\label{fig:007}
\end{figure*}
\subsection{Blockchain Facet:}
To answer the RQ4, this facet classified the publications along the Blockchain attributes like framework (Ethereum, Hyperledger or other frameworks), type of Blockchain (public. private, permissioned) and focus of research paper on  consensus algorithm.

The results of mapping process are summarized in Figure \ref{fig:007} in form of the bubble plot to show the frequencies. This visualization gives a quick overview of publications for each category.

\section{Analysis}\label{sec:analysis}
This section discusses the study results and answers the research questions that we defined in Section \ref{sec:background}.\\ 

\textbf{RQ1: What are the problems between stakeholders in manufacturing industry?}\\\\
Manufacturing industry is facing security incidents due to the competition~\cite{Wang:2019:SSB:3302505.3310086}, data sharing with third companies~\cite{8678753}, operational inefficiencies, losses and costs~\cite{8626103}. Issue of limited trust is one of the complications in the industry~\cite{Geiger:2019:PTD:3297280.3297546,Innerbichler:2018:FBA:3211933.3211953,8704309}. In~\cite{Pinheiro2019331} authors illustrate the dependency of industrial companies on Trusted-Third-Party (TTP). It caused by the closed source code of programs, used in manufacturing industry~\cite{8678753}. Moreover it is necessary to differentiating between original part or counterfeit products~\cite{Holland2018,BANERJEE201869}. The outsourcing of production orders~\cite{Baumung2019456} leads to limited flexibility  and considerable organizational effort.
\\ \\
\textbf{RQ2: What are the data to secure in manufacturing process?} \\\\
For some businesses the data exchange is a key success factor, so we found several type of data that should be protected between stakeholders: 
\begin{itemize}
\item Computer-Aided Design (CAD) file for design and technical documentation~\cite{Holland2018},~\cite{Papakostas2019}
\item Material specification~\cite{Mondragon20181300,Papakostas2019}
\item Order details~\cite{Baumung2019456} and product recipe~\cite{WESTERKAMP2019}
\item Machine data~\cite{Geiger:2019:PTD:3297280.3297546,ANGRISH20181180} (measurement data~\cite{Wang:2019:SSB:3302505.3310086} and configuration\cite{Mondragon20181300})
\item Process values~\cite{Geiger:2019:PTD:3297280.3297546,MANDOLLA2019134} and process state~\cite{LI2018133,8704309,8621042}
\end{itemize}

Selection of data, which should be protected, depends on specific scenario and application area of Blockchain. \\ 
\textbf{RQ3: What are the use cases of Blockchain technology for manufacturing industry?} \\ \\
There are various use cases of Blockchain technology in the industrial manufacturing. In this study we identified 5 papers (17\% of all papers) describing use case "Secure transfer of order data" in form of solution proposal. All of this papers provide prototypical implementation. Only 4 papers (14\% ) are relevant for storing of product data in distributed ledger. According to our findings, significant part of papers (around 43\%) illustrate solution proposal for supply chain and process traceability and 9 of them (32\%) demonstrate implementation examples. Fraud prevention and IP Protection are considered in 2 papers (around 7\%) and both of these papers implement this use case. The last use case "IoT, Automation" are described in 5 papers and covers equally all types of research.
The application of Blockchain is not limited to the use cases above. \\ \\
\textbf{RQ4: What Blockchain frameworks are suitable for the scenario "Assignment of production orders to an external manufacturer"?} \\ \\
This scenario is implemented mostly on Ethereum Blockchain (in 4 of all papers)~\cite{Baumung2018135,Baumung2019456,Papakostas2019,ANGRISH20181180}. In~\cite{Baumung2019456} was illustrated an example of using Hyperledger Fabric and in~\cite{LI2018133} was used Multichain. Public Blockchain was used in 2 papers~\cite{Baumung2018135},~\cite{Baumung2019456}, in 1 research work the authors used the private Blockchain~\cite{Papakostas2019} and in 4 papers was chosen consortium or permissioned Blockchain~\cite{LI2018133,Baumung2018135,Baumung2019456,ANGRISH20181180}. Some papers describe several types of Blockchain in the case use case. It means, that this use case is possible to implement based on different Blockchain networks and doesn't require a specific framework.  
\section{Conclusion}\label{sec:conlcusion}
In the coming years it is expected, that the manufacturing sector will benefit from the use of Blockchain technology. In order to identify opportunities for integration of Blockchain in industrial processes, this research was made in form of systematic mapping study. After conducting the SMS and analysing the literature, a total of 28 primary papers were extracted from 4 different scientific databases, published mainly in journals and conference proceedings and classified into different facets. We have covered the time period of 2017-2019 and have classified the papers under different dimensions. We grouped these issues into five use cases, namely, secure transfer of order data, product data storing, supply chain and process traceability, fraud prevention and IP protection, IoT and automation. 

In this study we found that the majority of papers describe the case "supply chain and process traceability" as solution proposal. There are essential less findings regarding assignment of production orders to an external manufacturer. It demonstrates the relative lack of research on this scenario, that requires effort lot more research. As found out, the last use case can be implemented using different frameworks. For example, this case could be implemented base on permissioned Blockchain using Hyperledger Fabric\footnote{\url{www.hyperledger.org}}. However, it is required to evaluate all frameworks to select the most suitable solution for the use case. 

The result of this mapping study can be applied only on the selected research databases and may help the researchers to get an overview of the status of Blockchain in the manufacturing industry and highlight the research gaps. 

Our line of future research aims precisely to implement our own prototype of use case "Assignment of production orders to an external manufacturer" to demonstrate the benefits of using Blockchain in factory automation. Additionally, we are going to extend the literature survey to include other databases like SpringerLink and we will apply snowballing to assure that the search is as comprehensive as possible. 

\bibliographystyle{IEEEtran}

\end{document}